\documentclass[reprint,amsmath,amssymb,aps]{revtex4-2}

\usepackage{graphicx}
\usepackage{dcolumn}
\usepackage{bm}
\usepackage{xcolor}
\usepackage{soul}

\begin{document}

\preprint{APS/123-QED}

\title{Effect of interface on magnetic exchange coupling in Co/Ru/Co trilayer: from \emph{ab initio} simulations to micromagnetics}

\author{Sergiu Arapan}
 \affiliation{ VSB - Technical University of Ostrava}
 \email{sergiu.arapan@gmail.com}
 \author{Jan Priessnitz}
\affiliation{ VSB - Technical University of Ostrava}
\author{Alexander Kovacs}%
\affiliation{%
 Universit\"at f\"ur Weiterbildung Krems
}%
\author{Harald Oezelt}
\affiliation{%
 Universit\"at f\"ur Weiterbildung Krems
}%
\author{David Böhm}
\affiliation{%
 Universit\"at f\"ur Weiterbildung Krems
}%
\author{Markus Gusenbauer}
\affiliation{%
 Universit\"at f\"ur Weiterbildung Krems
}%
\author{Thomas Schrefl}
\affiliation{%
 Universit\"at f\"ur Weiterbildung Krems
}%

\author{Dominik Legut}
\affiliation{ VSB - Technical University of Ostrava}

\begin{abstract}
Interfaces play a substantial role for the functional properties of structured magnetic materials and magnetic multilayers. Modeling the functional behavior of magnetic materials requires the treatment of the relevant phenomena at the device level. Properties predicted from the electronic structure and spin dynamics at the atomistic level have to be properly transferred into a continuum level treatment. In this work we show how Co/Ru/Co three layers can be simulated with the continuum theory of micromagnetism, with interface coupling energies and bulk intrinsic properties properly derived from the results of \emph{ab initio} and spin dynamics simulations at different temperatures.
\end{abstract}

\maketitle

\section{\label{sec:intro}Introduction\protect}
 
In magnetic materials, spins/magnetic moments are coupled. This coupling is described by a quantum mechanical effect called exchange. It is the main physical effect/reason for the occurrence of various magnetic configurations with broken symmetry, such as the ferromagnetic, anti-ferromagnetic, ferrimagnetic phases, etc. The exchange energy is the energy that misaligned spins/magnetic moments/electrons have to pay and reason that in equilibrium more aligned ones are energetically favored. How strong spins/magnetic moments are coupled is determined by either the exchange integral $J_\mathrm{ij}$ , depending on the overlap of wave functions of two unpaired electrons or the more classical exchange constant $A$, which is the prefactor for the following exchange energy density $e_\mathrm{ex}= A|\nabla \bm{m}|^2$, where $\bm{m}$ denotes the magnetization direction~\cite{coey_2010}. Magnetic phenomena are studied through their dynamic behavior, and the systems are, usually, too large to be investigated at the atomistic level. Thus, the magnetic properties are predicted within the micromagnetic description. In the framework of micromagnetism, all relevant energies are expressed in terms of the unit magnetization vector which is a continuous function of space and time $\bm{m}(t,\bm{x})$~\cite{Fidler_2000}. The micromagnetic approach is based on the Landau-Lifshitz-Gilbert (LLG) equation for $\bm{m}(t,\bm{x})$, and its solutions require the knowledge of appropriate structural and micromagnetic properties.
One approach is to determine these parameters from the intrinsic magnetic properties at the atomic level, by mapping  the results of the electronic structure calculations to a Heisenberg model. A shortcoming of this method is that the intrinsic magnetic properties are derived for ideal bulk systems, while the real materials are never perfect. Another aspect is that classical micromagnetism does not treat thermal fluctuations explicitly. The intrinsic magnetic properties are assumed to depend on temperature. The magnetic moment per unit volume is replaced by its average temperature dependent value which results in the temperature dependent spontaneous magnetization $ M_\mathrm{s}(T)$. Similarly, the magnetocrystalline anisotropy constant $K(T)$ and the exchange constant $A(T)$ are temperature dependent. They are normally measured at temperature $T$, at which the specimen is observed. 

 In this work, we show how these temperature-dependent parameters can be derived for a system with interfaces from \emph{ab initio} simulations and atomistic spin dynamics (ASD). The interfaces in magnetic structured materials are responsible for the occurrence of many physical phenomena of relevance to modern technological applications~\cite{RevModPhys.89.025006}. There are a variety of interesting effects that take place in trilayers or multilayers of ferromagnetic materials separated by thin layers of nonmagnetic metals. One observation, directly related to the exchange coupling between the ferromagnetic layers, is the oscillation of the interlayer exchange coupling with the thickness of the nonmagnetic layer~\cite{parkin1990oscillations,SLONCZEWSKI199513,STILES1999322}. As an example, we study the coupling of two Co layers in a Co/Ru/Co trilayer. Depending on the Ru layer thickness this coupling is either ferromagnetic or antiferromagnetic~\cite{mckinnon2021spacer,PhysRevApplied.22.024058}. For the micromagnetic simulations of the trilayer, we assume an interlayer exchange constant between two Co layers $J_\mathrm{Co,Co}(T)$, which is temperature dependent. Its value is derived by matching the interface spin configuration obtained from spin dynamic simulations and micromagnetic theory.

\begin{figure}[h] 
	\includegraphics[width=\columnwidth]{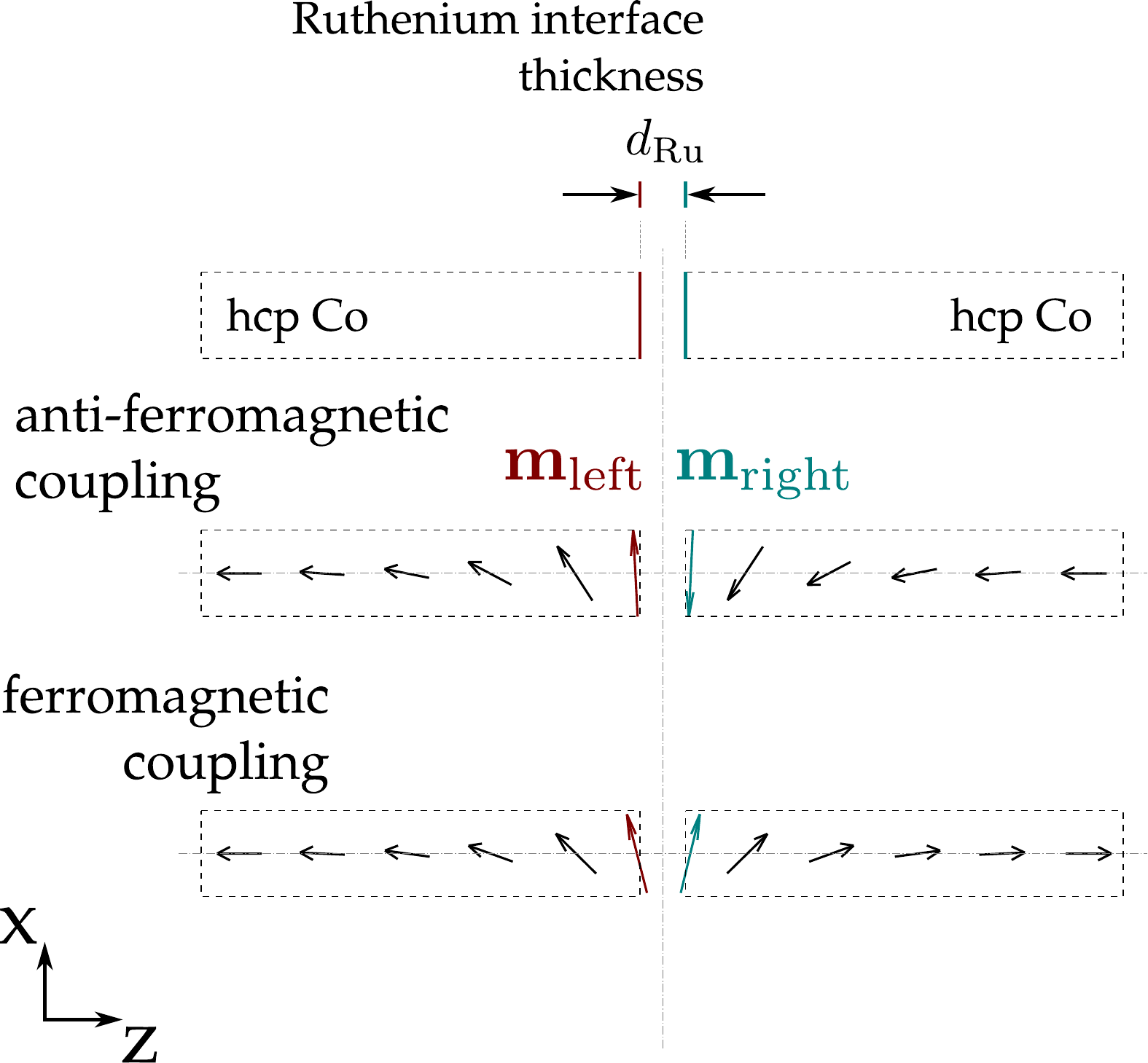}
	\caption{\label{fig:spinchain}Simulation of magnetic moment configurations for domain walls across a Ru layer in hcp Co. Top: Schematics of the Co/Ru/Co trilayer. Center: Expected magnetization profile of a domain wall for anti-ferromagnetic coupling across the Ru layer. Bottom: Expected magnetization profile for ferromagnetic coupling across the Ru layer.}
\end{figure}

The main subject of our study is the spin configuration of a domain wall at the Co$|$Ru interface. The dynamical behavior of the domain wall structure is simulated as a function of temperature using both ASD simulations and micromagnetic simulations. Fig.~\ref{fig:spinchain} shows the expected magnetization distribution across the Ru layer in hexagonal close packed (hcp) Co.

This paper is structured as follows: 

First, we present the general methodology (\ref{sec:method}). We summarize the procedure (\ref{sec:overview}) applied to obtain the intrinsic material properties for micromagnetic simulations of Co/Ru/Co trilayers. We show how we derived the exchange interaction energy per atom $J_\mathrm{int}$ across the Ru layer (\ref{sec:abinit}) using \emph{ab initio} simulations. This is an input parameter for the ASD simulations (\ref{sec:spindynintro}) in detail. From the results of the ASD simulations, we derived the input parameters of micromagnetic simulations (\ref{sec:micromagnetic}). 

Second, we present the results (\ref{sec:results}). We show the temperature-dependent domain wall profiles in Co/Ru/Co trilayers computed with the ASD simulations for varying Ru thickness (\ref{sec:spindynamicprofiles}). We derive the temperature dependent intrinsic magnetic properties for micromagnetic simulations from the spin dynamics results (\ref{sec:microproperties}). We use the temperature dependent intrinsic magnetic properties in micromagnetic simulations of the domain wall profiles for different Ru thickness (\ref{sec:micromagneticprofiles}). 

Finally, we draw conclusions (\ref{sec:conclusion}).

\section{\label{sec:method}Methodology}
\subsection{\label{sec:overview}Overview}
\begin{figure}[h] 
	\includegraphics[width=\columnwidth]{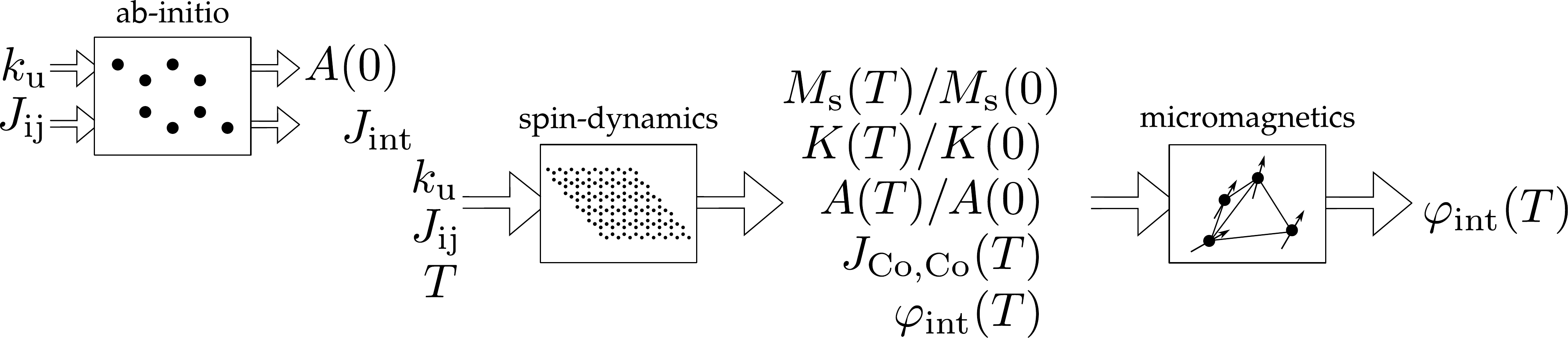}
	\caption{\label{fig:passing}Parameter passing between \emph{ab initio} calculations, spin dynamics simulations and micromagnetic calculations, see~(\ref{sec:overview}) for more details.}
\end{figure}

For the micromagnetic simulations of the domain wall profiles of Co/Ru/Co trilayers as a function of temperature, we need the temperature dependent exchange constant $A(T)$ of Co, the temperature dependent anisotropy constant $K(T)$ of Co, and the interlayer exchange constant $J_\mathrm{Co,Co}(T)$ between the two Co layers across the Ru interface. The latter will depend on the Ru thickness. In order to derive these parameters we apply the following procedure.
\begin{enumerate}
\item \emph{Ab initio} simulations \\
\emph{Ab initio} electronic structure calculations can predict the $T=0 K$ intrinsic magnetic properties such as the total magnetic moment of the system and magnetic (both spin and orbital) moments of individual atoms, the exchange interaction between magnetic moments at various atomic sites, and the magneto-crystalline anisotropy energy. In this work we calculate the exchange energy between atom pairs across the Ru interface $J_\mathrm{int}$. For the exchange interaction energy between pairs of magnetic moments $J_{ij}$ in bulk Co we use the values previously computed by Turek et al. \cite{Turek2003}, and the anisotropy per atom $k_u$ is taken from Moreno et al.~\cite{Moreno2016}. 

\item Atomistic spin dynamic simulations \\
ASD simulations use the exchange energies $J_{ij}$ and $J_\mathrm{int}$, and the anisotropy energy $k_\mathrm{u}$ per atom as input. These simulations solve the LLG equation for classical spins at the fixed atom sites. Temperature is included through a thermal fluctuation field whose strength is derived from the fluctuation-dissipation theorem. 
We show that the temperature dependent anisotropy constant $K(T)/K_0$ scales like $\left(M_\mathrm{s}(T)/M_\mathrm{s}(0)\right)^3$. Using computed domain wall profiles for bulk Co, we obtain the temperature dependent $A(T)$ and show that $A(T)/A(0)$ scales like $\left(M_\mathrm{s}(T)/M_\mathrm{s}(0)\right)^2$.  Spin dynamic simulations of magnetization profiles of Co/Ru/Co trilayers show a jump in the average magnetic moment across the Ru layer. We define the angle $\varphi_\mathrm{int}$ between the average of magnetic moments of Co touching the left side of the Ru layer and the average of the magnetic moments of the Co atoms touching the right side of the Ru layer.

\item Micromagnetism \\
In classical micromagnetism, the magnetic state is described by the vector field $\bm{m}(t,\bm{x})$. Since we treat the exchange coupling of ferromagnetic layers across interfaces, the magnetization $\bm{m}(t,\bm{x})$ is not continuous across the interface \cite{Skomski2001}. We use micromagnetic theory to derive the temperature dependent exchange coupling $J_\mathrm{Co,Co}(T)$ across the Ru layer and magnetization profiles in Co/Ru/Co layers from $K(T)$, $A(T)$, and $\varphi_\mathrm{int}$ obtained from the ASD simulations. 
 
\item Validation \\
We compare the temperature dependent magnetization profiles in Co/Ru/Co layers computed with spin dynamics simulations and micromagnetic simulations.

\item \emph{Nota Bene} \\
Here we should mention that, in this work, we mainly focus on the exchange between ferromagnetic layers and how it is affected by the Ru spacer. At this stage we do not take into account the changes in the spin magnetic moment of Co atoms at the Co$|$Ru interface, as well as the variation of the magneto-crystalline anisotropy of the Co/Ru/Co systems with the width of the Ru spacer. We also do not consider the change of the magneto-crystalline anisotropy due to structural phase transition at elevated temperatures. 
\end{enumerate}
The above scheme for parameter passing from the \emph{ab initio} calculations to the ASD, and micromagnetic simulations is schematically shown in Fig.~\ref{fig:passing}.

\subsection{\label{sec:abinit}\emph{Ab initio}  calculations}

\begin{figure}[h] 
	\includegraphics[width=\columnwidth]{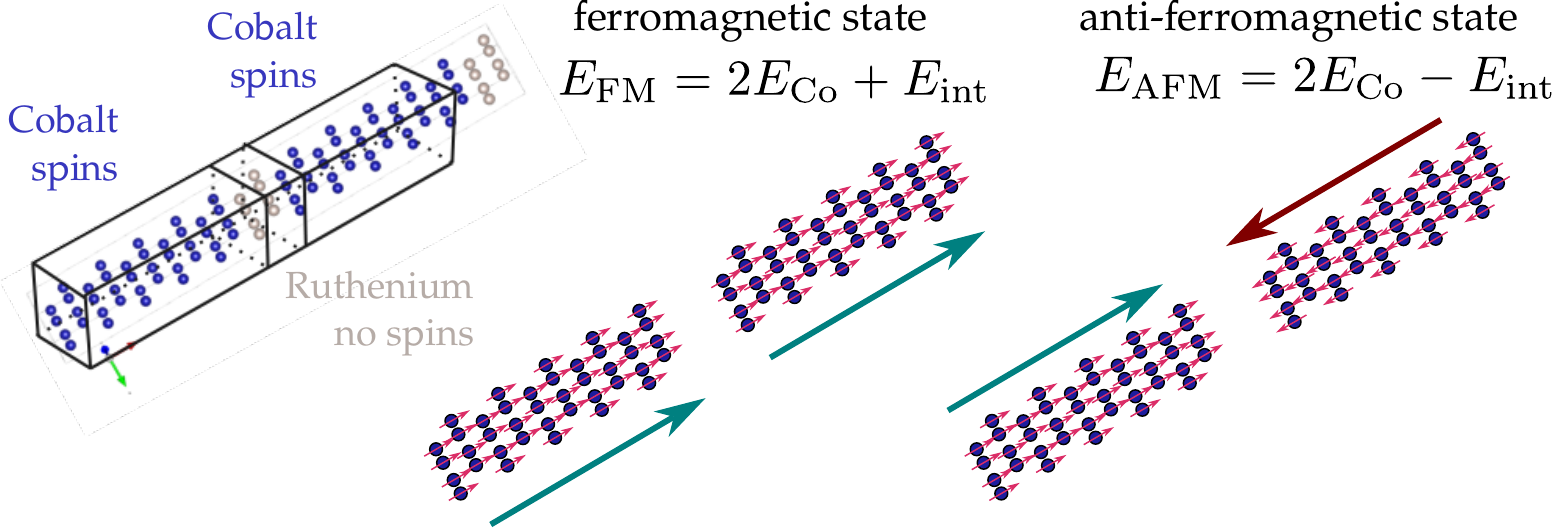}
	\caption{\label{fig:dft-afm-fm-ei-comp}Sketch of a Co/Ru/Co slab. Both Co bulk parts are configured in two different states i) into a ferromagnetic (FM) state and ii) into an anti-ferromagnetic (AFM) state. The calculated total energies $E_\mathrm{AFM/FM}$ of the AFM/FM configurations can be factorized in the contributions from the two bulk-like Co slabs $E_\mathrm{Co}$, and exchange interaction energy at the interface $E_\mathrm{int}$.}
\end{figure}

    Electronic structure calculations of Co/Ru/Co systems were done with the Vienna Ab initio Simulation Package (VASP) software (version 6.2.1). VASP is a plane-wave basis set implementation~\cite{VASP1,VASP2} of the Density Functional Theory within the Projector Augmented Wave (PAW) approximation.~\cite{PAW} Calculations were performed using the generalized gradient approximation of Perdew, Burke, and Ernzerhof (PBE)~\cite{PBE} to the exchange-correlation part of the energy functional. We used PAW PBE potentials version 5.4 with 15 and 14 valence electrons for Co and Ru, respectively. The size of the plane-wave basis set was determined by an energy cut-off of 440 eV. The reciprocal space was sampled with a uniform mesh of k points~\cite{KMESH} with a separation between the k points of 0.1 \AA$^{-1}$. For total energy summation, we used the Methfessel-Paxton method~\cite{ISMEAR}, with a smearing width of 0.025 eV. Electronic convergence was set to 10$^{-7}$eV, and geometry optimization was carried out until the norms of all Hellman-Feynman forces~\cite{FORCES} were less than 10$^{-3}$eV/\AA. Equilibrium structure parameters were obtained by performing a set of geometric optimizations (ion positions and cell shape) at volumes within a 5$~\%$ range about equilibrium and fitting the energy vs volume data, E = E(V), to the Rose-Vinet equation of state.~\cite{RV-EOS} 
    Co($d_\mathrm{Co}$)/Ru($d_\mathrm{Ru}$)/ Co($d_\mathrm{Co}$) multilayers were modeled as periodic supercells of the hcp unit cell, with the $c-$ axis oriented along the $x-$direction. A supercell comprises two Co layers, with layer width 0.8 nm $\leq d_\mathrm{Co} \leq$ 2.6 nm  (from 4 to 13 Co atomic layers) each, separated by a few Ru atomic layers (1, 2, and 3) as shown in Fig.~\ref{fig:dft-afm-fm-ei-comp}.

From the calculated total energies, we can now calculate an effective exchange interaction energy per atom as following:
\begin{equation}
	E_\mathrm{int}=(E_\mathrm{FM}-E_\mathrm{AFM})/2 ,
    \label{eq_Eint} 
\end{equation}
where $E_\mathrm{AFM/FM}$ are the total energies of the AFM/FM configurations. $E_\mathrm{int}$ represents the exchange interaction of the left bulk with the right bulk across the Ru interface.

According to the Heisenberg model, we will add $E_\mathrm{int}$ to the total energy when the spins across interface are oriented in parallel, and subtract $E_\mathrm{int}$ if the spins are oriented anti-parallel. Thus, $E_\mathrm{int}$ can be calculated by subtracting the total energies $E_\mathrm{FM}$ and $E_\mathrm{AFM}$. With $N_\mathrm{int}$ Co atoms in the layer adjacent to the interface, the effective Co-Co exchange coupling per pair across the Ru spacer is simply given by:

\begin{equation}
	J_\mathrm{int}=E_\mathrm{int} / N_\mathrm{int}
    \label{}
\end{equation}

\begin{figure}[h] 
	\includegraphics[width=\columnwidth]{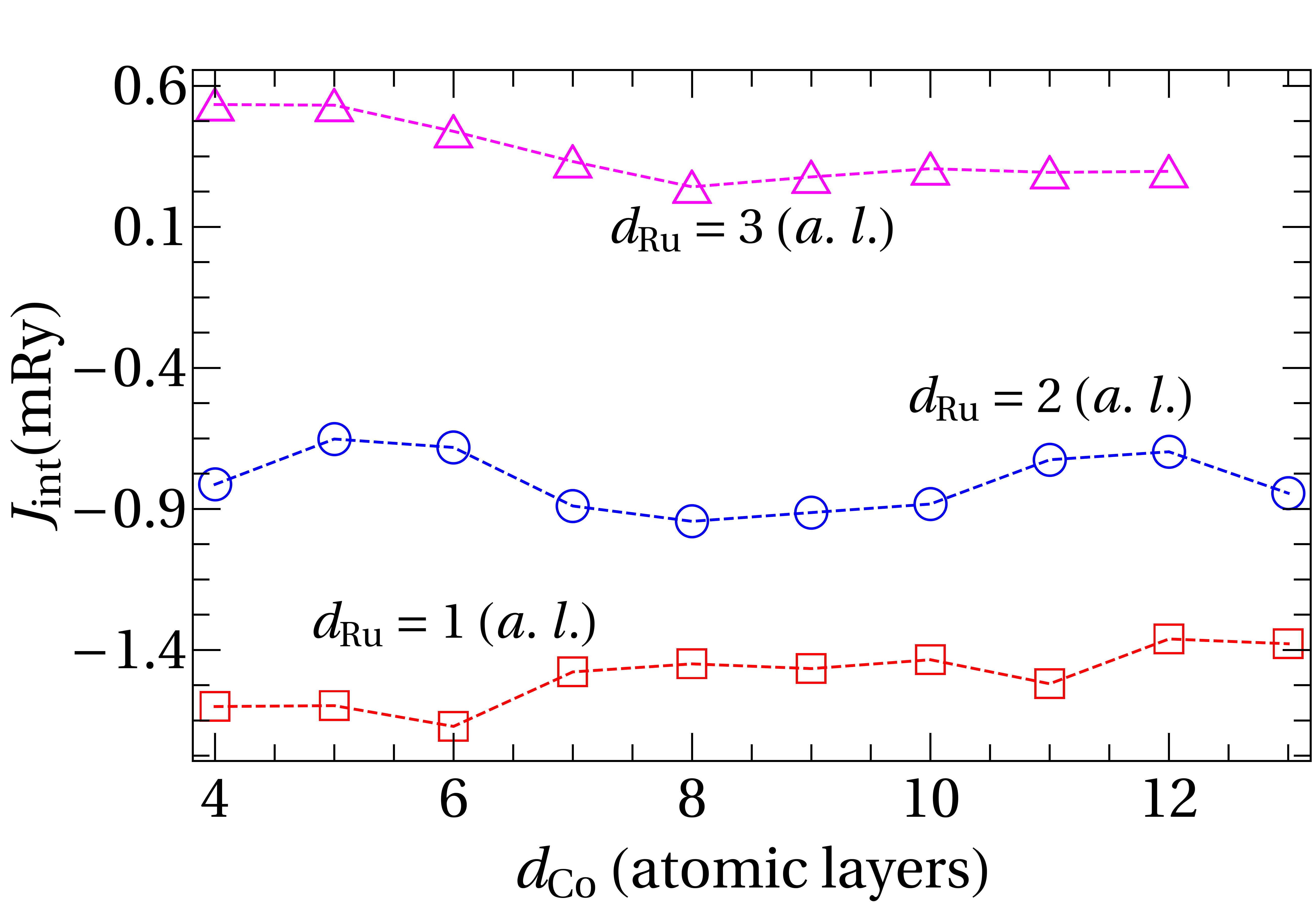}
	\caption{\label{fig:Jeff_dCo_dRu} Effective Heisenberg interactions $J_{int}$ between Co atoms separated by a Ru spacer as a function of the width of Co layers $d_\mathrm{Co}$ and Ru spacer $d_\mathrm{Ru}$.}
\end{figure}

 The resulting coupling energies for 1, 2 and 3 monolayers of Ru as a function of the width of Co layers are shown in Fig.\ref{fig:Jeff_dCo_dRu}. For the ASD calculations, we have chosen the following values for the effective coupling between Co pairs separated by a Ru spacer: $J_\mathrm{int,1} = -1.376\mathrm{~mRy}\;(-3\times 10^{-21}$~J) ), $J_\mathrm{int,2} = -0.7301\mathrm{~mRy}\;(-1.59 \times 10^{-21}$~J) and $J_\mathrm{int,3} = 0.3011\mathrm{~mRy}\;(0.656\times 10^{-21}$~J) for a Ru spacer of 1, 2, and 3 atomic layers, respectively.

Following the work of  Moreno et al. \cite{Moreno2016}, we compute the zero temperature exchange constant of bulk hcp Co parallel to the $c$-axis using
\begin{equation}
	A(0) = \frac{1}{4 V_\mathrm{at}} \sum_{i} J_{0 i} \cdot (r_i - r_0)^2,
    \label{eq:A0}
\end{equation}
where $i$ iterates over all neighboring atoms of 0-th atom, $J_{0i}$ is the exchange interaction energy, $r_i$ is the atom positions along the $c$-axis, and $V_\mathrm{at} = 1.1 \times 10^{-29}$~m$^3$ is the volume per atom. This gives $A(0) = 52.78$~pJ/m.

\subsection{\label{sec:spindynintro}Spin dynamic simulations}

We use the classical spin dynamics method to bridge the gap between density functional theory, which calculates supercells of up to hundreds of atoms, and micromagnetics, which simulates systems on the micrometer scale containing billions of atoms, albeit on the continuum level. Our ASD usage lies somewhere in the middle, simulating around 1 million atoms. This is necessary to capture the collective phenomena that arise due to finite temperature and determine the temperature dependence of the magnetic properties of the material, such as saturation magnetization $M_s(T)$, magnetocrystalline anisotropy $K(T)$, exchange stiffness $A(T)$ and, in turn, the domain wall width $\delta_{DW}(T)$.

The simulated system is described by the following Hamiltonian:
\begin{equation}
	\mathcal{H} = - \sum_{i, j} J_{ij} \vec{S}_i \cdot \vec{S}_j - \sum_{i} k_\mathrm{u} (\vec{S}_i \cdot \vec{e}_k)^2- \sum_{i} \vec{S}_i \cdot \vec{B}_{ext},
    \label{eq:H_ASD}
\end{equation}
where $J_{ij}$ are the exchange interaction energies between pairs of magnetic moments i and j, $k_u$ is the magnetocrystalline anisotropy energy, $\vec{e}_k$ is the easy-axis direction ($\vec{e}_k = (0, 0, 1)$ throughout this work), $\vec{S}_i$ is the normalized moment vector at $i$-th site ($|\vec{S}| = 1$), and $\vec{B}_{ext}$ is the external magnetic field vector. Note that each pair is counted twice ($J_{ij}$ is halved to account for double counting).

The exchange interaction energies $J_{ij}$ are taken from ab initio calculations by Turek et al.~\cite{Turek2003}. 38 nearest unique exchange interaction pairs were taken into account, corresponding to a distance cutoff of about 1~nm.  
For the anisotropy energy per atom we used the value $k_\mathrm{u} = 5.83 \cdot 10^{-24}~\mathrm{J/atom}$~\cite{Moreno2016}, corresponding to a macroscopic magnetocrystalline anisotropy constant of $K(0) = 0.547~\mathrm{MJ/m^3}$. These values yield a $T=0 K$ domain wall width of $\delta_{DW} = \pi \sqrt{A/K} =  31.35~\mathrm{nm}$.

In this work, we simplify the actual Co/Ru/Co interface by replacing the influence of the Ru atoms with an effective exchange interaction $J_\mathrm{int}$ (Section~\ref{sec:abinit}). Apart from this effective exchange interaction, there are no other interactions between atoms of the two bulks. Thus we write for the Hamiltonian for the Co/Ru/Co trilayer:

\begin{eqnarray}
    \sum_{i, j} J_{ij} \vec{S}_i \cdot \vec{S}_j =
	\sum_{i, j \in C_{\mathrm{BL}}} J(\vec{r}_{ij}) \vec{S}_i \cdot \vec{S}_j~+ \nonumber\\
	\sum_{i, j \in C_{\mathrm{BR}}} J(\vec{r}_{ij}) \vec{S}_i \cdot \vec{S}_j~+ \nonumber\\
	\sum_{i \in C_{\mathrm{IL}}, j \in C_{\mathrm{IR}}} J_{\mathrm{int}} \vec{S}_i \cdot \vec{S}_j \delta(\vec{r}_i - \vec{r}_j - \vec{r}_{\mathrm{int}})~+ \nonumber\\
	\sum_{i \in C_{\mathrm{BL}}, j \in C_{\mathrm{BR}}} (-1) \cdot J(\vec{r}_{ij} - \vec{l}_z) \vec{S}_i \cdot \vec{S}_j,
      \label{eq:Heis_sum}
\end{eqnarray}
where $C_{BL}$, $C_{BR}$ are sets containing spins in the left and right bulk, respectively, $C_{IL}$ and $C_{IR}$ are sets of spins on the left and right side of the interface and $J_{\mathrm{int}}$ is the specific energy of exchange interaction across the interface calculated from bilinear coupling in section \ref{sec:abinit}. $\vec{r}_{\mathrm{int}}$ is the relative position of the nearest neighbors across the interface and $\vec{l}_z$ is the size of the simulation domain in the z-direction. Note that $C_{IL} \subset C_{BL}$, $C_{IR} \subset C_{BR}$ and $C_{BL} \cup C_{BR}$ contains all spins.

The first two terms describe the exchange interactions inside the left and right Co bulk. The third term stands for the interaction across the Ru interface. The fourth term stands for the antiperiodic boundary condition in the z-direction, which helps to pin the domain wall in the middle of the simulation domain. Periodic boundary conditions in the (x,y)-direction are also employed in this model, but not explicitly mentioned in the Hamiltonian above.

The atomistic ASD approach is based on the LLG equation for individual spins:

\begin{equation}
    \frac{\mathrm{d} \vec{S}_i}{\mathrm{d} t} = - \frac{\gamma}{1 + \alpha^2} \vec{S}_i \times \vec{H}^{\mathrm{eff}}_{i} - \frac{\gamma \alpha}{1 + \alpha^2} \vec{S}_i \times (\vec{S}_i \times \vec{H}^{\mathrm{eff}}_i),
    \label{eq:ASD_LLG}
\end{equation}
where $\gamma$ is the gyromagnetic ratio, $\alpha$ is the Gilbert damping parameter and $\vec{H}^{\mathrm{eff}}_i$ is the effective field experienced by $i$-th spin. To introduce temperature $T$, a stochastic Gaussian field $\vec{b}_i(t)$ with variance of $\sqrt{\frac{2 \alpha k_B T}{\Delta t \gamma \mu_B}}$ is added to the effective field:
\begin{equation}
   \vec{H}^{\mathrm{eff}}_i = - \frac{\mathrm{d}\mathcal{H}}{\mathrm{d}\vec{S}_i} + \vec{b}_i(t),
   \label{eq:eff_field}
\end{equation}
where $\Delta t$ is the simulation timestep. An ASD simulation is then formally just an integration of the LLG equation. In this work, Heun's method is used for numerical integration with a timestep $\Delta t$ small enough to keep the integration error negligible. The damping rate was chosen as $\alpha = 0.05$ based on the speed of convergence. The choice of damping parameter does not influence the domain wall width, since it is measured in the equilibrated state.

The cross section of the simulation domain is chosen to be large enough such that the thermal noise does not destroy the domain wall. The length is chosen such that the entire domain wall profile is simulated. In this work, the simulation domain consists of $50 \times 50 \times 240$ unit cells for hcp Co. Each unit cell contains 2 Co atoms at fractional coordinates $(0, 0, 0)$ and $(1/3, 1/3, 1/2)$, and the lattice parameters are $a = 2.47$ \AA~and $c/a = \sqrt{8/3}$. The simulation domain for the Co/Ru/Co systems is constructed by altering the original hcp Co simulation domain: first, a simulation domain of $60 \times 60 \times 240$ hcp Co unit cells is created, then $n$  central layers of atoms are removed to account for the Ru interface, and then the pairwise exchange interactions at the interface are modified.

All ASD calculations were carried out by our unpublished and in-house developed software. The main result of the ASD stage in this work is the magnetization profile along the z-direction $M(z) \in \mathbb{R}^3$, which is calculated as the average of all spins in the individual hcp layers. 

\subsection{\label{sec:micromagnetic}Micromagnetism}
Micromagnetism \cite{brown1963micromagnetics} is a continuum theory. The energy functional is expressed in terms of the unit magnetization vector $\mathbf{m}(\mathbf{x})$, which is a continuous function of position $\mathbf{x}$. We now derive the micromagnetic energy of the domain configurations shown in Fig.~\ref{fig:spinchain}. We assume translational symmetry in the $xy$ plane.  
In our approach we neglect magnetostatic interactions. There is no energy difference between a Bloch or a N\'{e}el wall \cite{hubert2008magnetic}. We restrict the vector $\mathbf{m}$ to lie in the $xz$ plane and describe its direction with the angle $\varphi$ from the postive $z$ axis. We start with the energy of the domain wall in bulk Co:
\begin{equation}
	\label{eq_wall}
	E(\varphi)  = \int_{a}^{b}  { \left( A\left(\frac{\mathrm{d} \varphi}{\mathrm{d} z}\right)^2 + K \sin^2 \varphi \right)   dz} 
\end{equation}
For a simple domain wall $\varphi (a) = \pi$ and $\varphi (b) = 0$. In order to express the energy of a domain wall across a Co/Ru/Co trilayer, we split the energy (\ref{eq_wall}) into two parts and add a surface energy term:
\begin{eqnarray}
	E_{\mathrm{wi}}(\varphi) & = &\;\;\int_{a}^{0}  { \left( A\left(\frac{\mathrm{d} \varphi}{\mathrm{d} x}\right)^2 + K \sin^2 \varphi \right)   dz} \nonumber \\ 
	&& + \int_{0}^{b}  { \left( A\left(\frac{\mathrm{d} \varphi}{\mathrm{d} z}\right)^2 + K \sin^2 \varphi \right)  dz}  \label{eq_total_energy} \\
	&& - J_\mathrm{Co,Co} \left( \mathbf{m}_\mathrm{left} \cdot \mathbf{m}_\mathrm{right} \right) \nonumber
\end{eqnarray}
The last term considers the coupling across the Ru layer. It is the interface energy expressed in terms of the magnetization vectors $\mathbf{m}_\mathrm{left}$ and $\mathbf{m}_\mathrm{right}$ left and right of the interface, respectively. We can write the interface energy in terms of the interface angle 
\begin{equation}
	E_\mathrm{int} = - J_\mathrm{Co,Co} \cos( \varphi_\mathrm{int} ) 
\end{equation}

with
\begin{equation}
	\label{eq:angleint}
	\varphi_\mathrm{int} = \varphi_\mathrm{left} - \varphi_\mathrm{right}.
\end{equation}

\begin{figure}[h]
	\centering	\includegraphics[width=0.7\columnwidth]{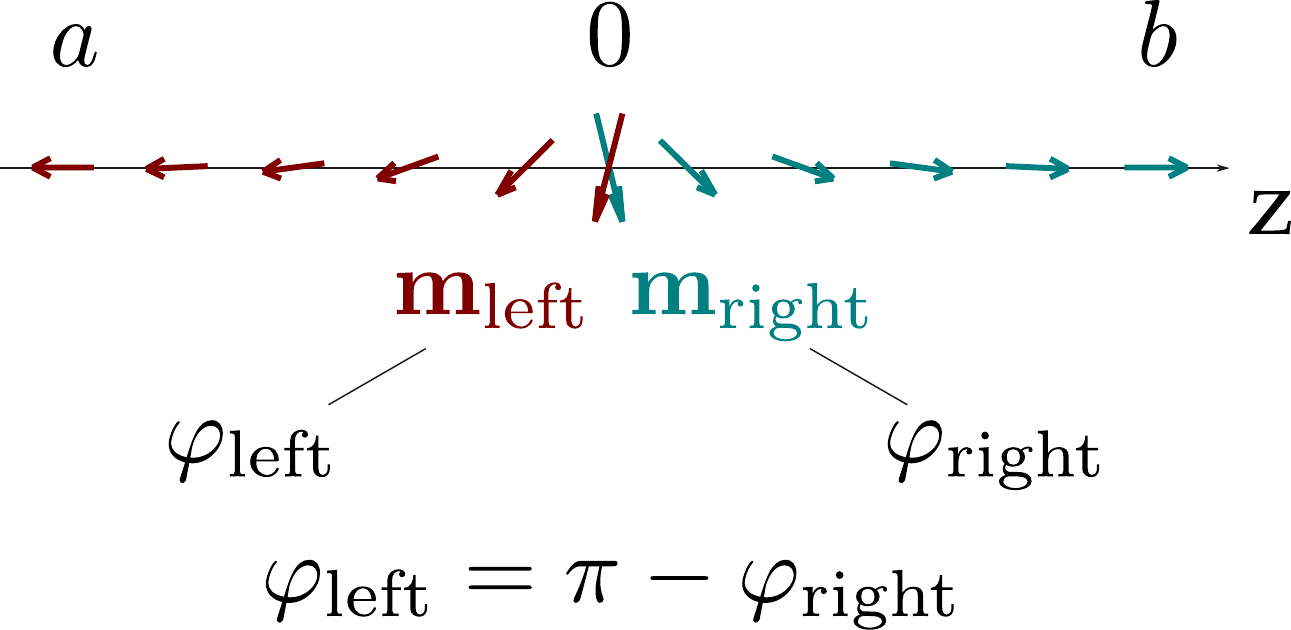}
	\caption{Magnetization configuration at a ferromagnetic interface.}\label{fig_wall}
\end{figure}

From the magnetization configuration in Fig.~\ref{fig_wall} we see that we can use symmetry to express $\varphi_\mathrm{right}$ in terms of $\varphi_\mathrm{left}$
\begin{equation}
	\label{eq:angleright}	
	\varphi_\mathrm{right} = \pi - \varphi_\mathrm{left} 
\end{equation} 
The interface energy is
\begin{equation}
	E_\mathrm{int} = - J_\mathrm{Co,Co} \cos(2\varphi_\mathrm{left} - \pi ). 
\end{equation}
or
\begin{equation}
	E_\mathrm{int} = J_\mathrm{Co,Co} \cos(2\varphi_\mathrm{left} ). 
\end{equation}

Owing to symmetry that the twist of the magnetization in the left and the right part  consumes the same amount of energy. 
Therefore, we can write for the total energy (see also Fig.~\ref{fig_wall}) as:
\begin{eqnarray}
	\label{eq:mumagenergy}
	E_{\mathrm{wi}}(\varphi) & = 
	& 2 \int_{a}^{0}  { \left( A\left(\frac{\mathrm{d} \varphi}{\mathrm{d} z}\right)^2 + K \sin^2 \varphi \right)  dz} \nonumber\\
	&& + J_\mathrm{Co,Co} \cos \left(2\varphi(0) \right)
\end{eqnarray}

For the functional 
\begin{equation}
	E(\varphi) = \int_a^0 \left( e \left( \varphi(z), \frac{\mathrm{d}\varphi(z)}{\mathrm{d}z} \right) \right)   + E_\mathrm{int}(\varphi(0)) 
\end{equation} 
the Euler Lagrange equation is
\begin{equation}
	\frac{\partial e}{\partial \varphi} - \frac{\mathrm{d}}{\mathrm{d}z}\left( \frac{\partial e}{\partial \frac{\mathrm{d}\varphi}{\mathrm{d}z}} \right)  =  0 \label{eq_euler1} 
\end{equation}
The surface energy $E_\mathrm{int}$ at $z=0$ leads the following boundary condition
\begin{equation}	\label{eq_mumag_boundary_condition}
	\frac{\mathrm{d} E_\mathrm{int}}{\mathrm{d} \varphi} + \frac{\partial e}{\partial \frac{\mathrm{d}\varphi}{\mathrm{d}z}} = 0.
\end{equation}
The Euler Lagrange equations in micromagnetics and the boundary conditions arising from surface energies are discussed in \cite{fruchart2011lecture} (see equations I.44 to I.47 in \cite{fruchart2011lecture}). 
For our problem we can write the following relations. 
\begin{eqnarray}
	e & = & 2 A\left(\frac{\mathrm{d} \varphi}{\mathrm{d} z}\right)^2 + 2 K \sin^2 \varphi \\
	E_\mathrm{int} & = & J_\mathrm{Co,Co} \cos \left(2\varphi(0) \right) 
\end{eqnarray}

\begin{eqnarray}
	\frac{\partial e}{\partial \varphi}	& = & 4 K \sin\varphi \cos\varphi \\
	\frac{\partial e}{\partial \frac{\mathrm{d}\varphi}{\mathrm{d}z}}  & = & 4 A \frac{\mathrm{d}\varphi}{\mathrm{d}z} \label{eq_dedphiprime} \\
	\frac{\mathrm{d}E_\mathrm{int}}{\mathrm{d} \varphi} & = & - 2 J_\mathrm{Co,Co} \sin \left(2\varphi(0) \right) \label{eq_boundary_diff}
\end{eqnarray}

With the above equations we rewrite (\ref{eq_euler1})
\begin{equation}
	4 K \sin\varphi \cos\varphi =  4 A \frac{\mathrm{d}^2\varphi}{\mathrm{d}z^2}
\end{equation}

\begin{equation}
	2 K \sin\varphi \cos\varphi =  2 A \frac{\mathrm{d^2}\varphi}{\mathrm{d}z^2}
\end{equation}

\begin{equation}
	\label{eq_euler1a}
	\frac{\mathrm{d} } {\mathrm{d}\varphi} \left(K \sin^2 \varphi \right) =  
	2 A \frac{\mathrm{d}^2\varphi}{\mathrm{d}z^2}
\end{equation}
We multiply (\ref{eq_euler1a}) with $\mathrm{d}\varphi/\mathrm{d}z$ and integrate  (see \cite{fruchart2011lecture} equation I.38)
\begin{equation}
	\label{eq_fruchard_i38}
	K \sin^2 \varphi(0) - K \sin^2 \varphi(z) = A\left(\left.\frac{\mathrm{d} \varphi}{\mathrm{d} z}\right|_0\right)^2 - A\left(\left.\frac{\mathrm{d} \varphi}{\mathrm{d} z}\right|_z\right)^2  
\end{equation}
At $z = -\infty$ the first term on the left hand side and the first term on the right hand side of (\ref{eq_fruchard_i38}) vanish as we are in the center of the magnet. Following the arguments in \cite{fruchart2011lecture}, we obtain
\begin{equation}
	\label{eq_ex_ani}
	K \sin^2 \varphi(z) = A\left(\frac{\mathrm{d} \varphi}{\mathrm{d} z}\right)^2
\end{equation}  

Let us now have a closer look at the boundary condition at $z = 0$. From (\ref{eq_mumag_boundary_condition}) we obtain with (\ref{eq_dedphiprime}) and (\ref{eq_boundary_diff}) 
\begin{equation}
	\label{eq_boundary}
	- 2 J_\mathrm{int} \sin \left(2\varphi(0) \right) + 4 A \left.\frac{\mathrm{d}\varphi}{\mathrm{d}z}\right|_0 = 0
\end{equation}
From (\ref{eq_ex_ani}) we get
\begin{equation}
	\label{eq_dphidz}
	\frac{\mathrm{d} \varphi}{\mathrm{d} z} = \pm\sqrt{\frac{K}{A}}\sin \varphi	
\end{equation}
In our setting $\frac{\mathrm{d} \varphi}{\mathrm{d} z} < 0$ (see Figure \ref{fig_wall}). Therefore we take the right hand side of (\ref{eq_dphidz}) with the minus sign and insert it into (\ref{eq_boundary}). At $z = 0$ we have
\begin{equation}
	- 2 J_\mathrm{Co,Co} \sin \left(2\varphi \right) - 4 \sqrt{AK} \sin \varphi = 0
\end{equation}
With $\sin(2\varphi) = 2 \sin(\varphi) \cos(\varphi)$ we obtain
\begin{equation}
	\label{eq:atinter}
	- 4 J_\mathrm{Co,Co} \sin \varphi \cos \varphi - 4 \sqrt{AK} \sin \varphi = 0
\end{equation}
We now evaluate equation (\ref{eq:atinter}) at $z=0$. Using $\varphi_\mathrm{left} = \varphi(0)$ we can write for $\varphi_\mathrm{left} \neq 0$  we obtain
\begin{equation}
	J_\mathrm{Co,Co} \cos \varphi_\mathrm{left} =  -\sqrt{AK}  
\end{equation}
and can express the angle of the magnetization at the interface in terms of the bulk temperature dependent properties $A(T)$ and $K(T)$ and the temperature dependent interface coupling constant $J_\mathrm{Co,Co}(T)$
\begin{equation}
	\label{eq:eint_thetaint}
	\cos \varphi_\mathrm{left} = -\frac{\sqrt{AK}}{J_\mathrm{Co,Co}}.  
\end{equation}

The jump of the $z$-component of the magnetization across the interface, $\Delta m_{z}$ was computed by averaging the spin dynamic results near the interface. Eq.~\ref{eq:eint_thetaint} can be expressed with that jump as
\begin{equation}
	\label{eq:JCoCodmz}
	J_\mathrm{Co,Co} = \pm \frac{\sqrt{AK}}{\Delta m_{z}/2}.
\end{equation}

In the results section (\ref{sec:microproperties}), we will use equation (\ref{eq:JCoCodmz}) to obtain the temperature dependent coupling constant $J_\mathrm{Co,Co}(T)$ from the interface angle $\varphi_\mathrm{int}$, which was computed by ASD simulations, and the bulk properties $A(T)$ and $K(T)$.

\begin{figure}[h]
	\centering
    \includegraphics[width=0.9\columnwidth]{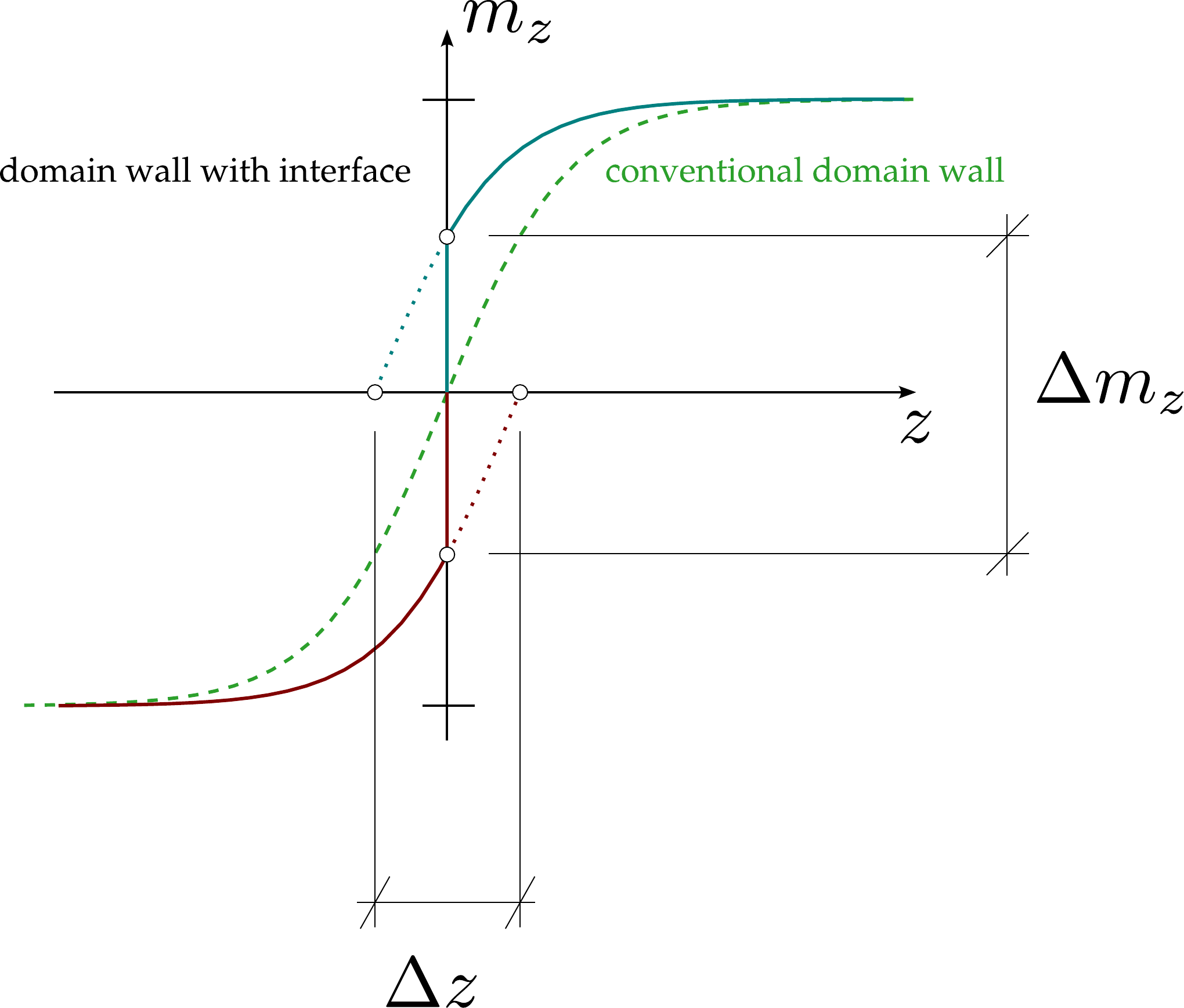}
	\caption{\label{fig:fig_wall}Illustration of a conventional domain wall and the magnetization distribution of a Co/Ru/Co trilayer with ferromagnetic coupling. The jump in the magnetization $\Delta m_{z}$ is used to compute the temperature dependent interface coupling constant for micromagnetics $J_\mathrm{Co,Co}$.}
\end{figure}

Fig.~\ref{fig:fig_wall} illustrates the jump of the magnetization across the interface for ferromagnetic coupling of the two Co layers in comparison to a conventional domain wall.

\section{\label{sec:results}Results}
\subsection{\label{sec:spindynamicprofiles}Domain wall profiles from ASD}

In the first stage, we use the ASD to calculate temperature-dependent domain wall properties in bulk hcp Co and compare with previously obtained results~\cite{Moreno2016}. The ASD calculation starts with an initial configuration where the left half of the spins is $\vec{S} = (0, 0, -1)$ and the right half is $\vec{S} = (0, 0, 1)$. The simulation then runs for 2,400,000 steps and the initial configuration relaxes into equilibrated domain walls.

The magnetization profile of the averaged configuration should then follow the shape of hyperbolic tangens for the $z$ component of the unit magnetization vector
\begin{equation}\label{eq:tanhfit}
m_z(z) = m_\mathrm{s} \tanh(\pi (z - z_0)/\delta_{\mathrm{DW}})
\end{equation}
where $m_\mathrm{s} = M_\mathrm{s}(T)/M_\mathrm{s}(0)$ is the reduced temperature dependent magnetization, $z_0$ is the domain wall center, and $\delta_{\mathrm{DW}}$ is the domain wall width. The simulated magnetization profile is then fitted into the above equation in order to obtain values for the free parameters $m_\mathrm{s}$, $z_0$ and $\delta_{\mathrm{DW}}$.
\begin{figure}[h]
	\includegraphics[width=\columnwidth]{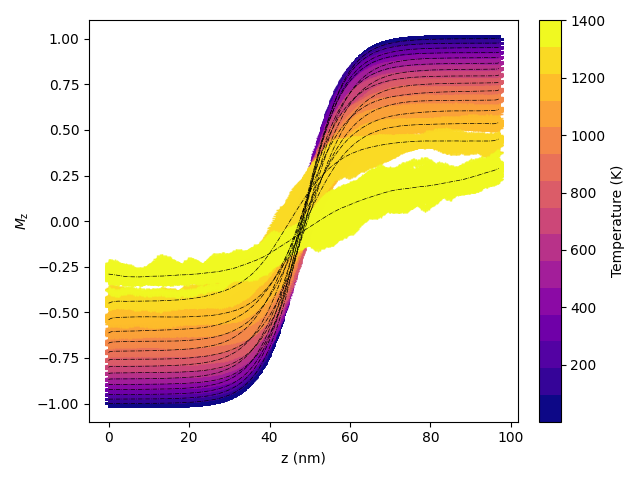}
	\caption{\label{fig:spin_dyn_walls_over_temp}Magnetization configurations after relaxation steps for different temperatures of hcp Co bulk. Colored regions show the averaged magnetization over one atomic Co layer of around 100 separate spin-dynamics simulations. The dot-dashed lines highlight the average of those profiles for each specific temperature.}
\end{figure}
\begin{figure}[h]
	\includegraphics[width=\columnwidth]{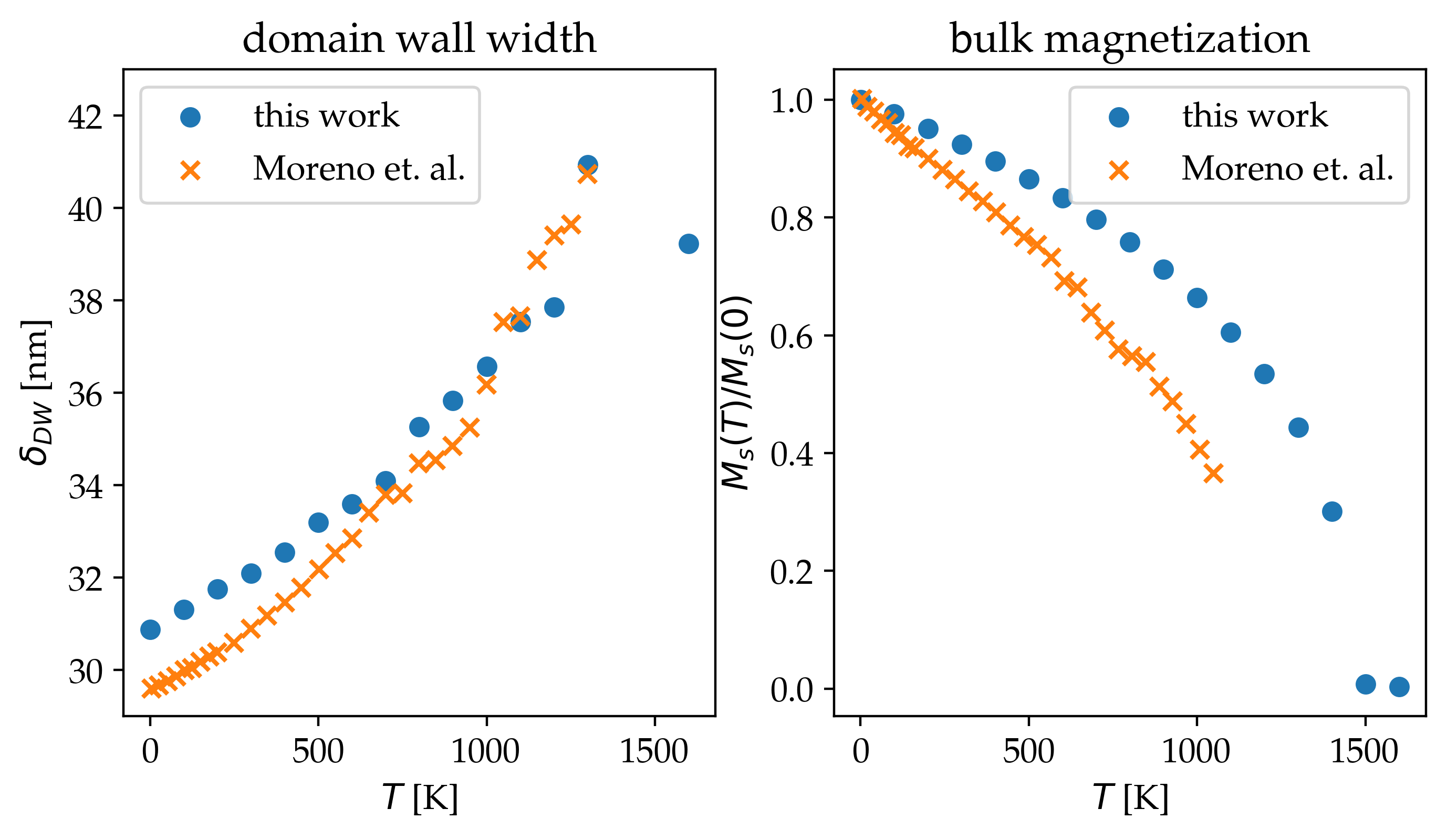}
	\caption{
		Temperature dependence of domain wall width and saturation magnetization in hcp Co extracted via fit of the Eq.~\ref{eq:tanhfit}. Results from Moreno et. al~\cite{Moreno2016} are shown for comparison.}
	\label{fig:ASD_hcpCo_width_satmag}
\end{figure}
The magnetization spin-dynamics profiles, their average profiles over temperature and the results of this fitting are shown in Fig.~\ref{fig:spin_dyn_walls_over_temp} and Fig.~\ref{fig:ASD_hcpCo_width_satmag}. In this work, the zero-temperature domain wall width is larger than in \cite{Moreno2016}, but increases more slowly with temperature. Also, the saturation magnetization decreases more slowly with temperature and indicates a Curie temperature of about $T_C \approx 1500~\mathrm{K}$, compared to $T_C = 1300~\mathrm{K}$ in \cite{Moreno2016}. This is caused by a different input set of exchange interactions.

\begin{figure}[h]
	\includegraphics[width=\columnwidth]{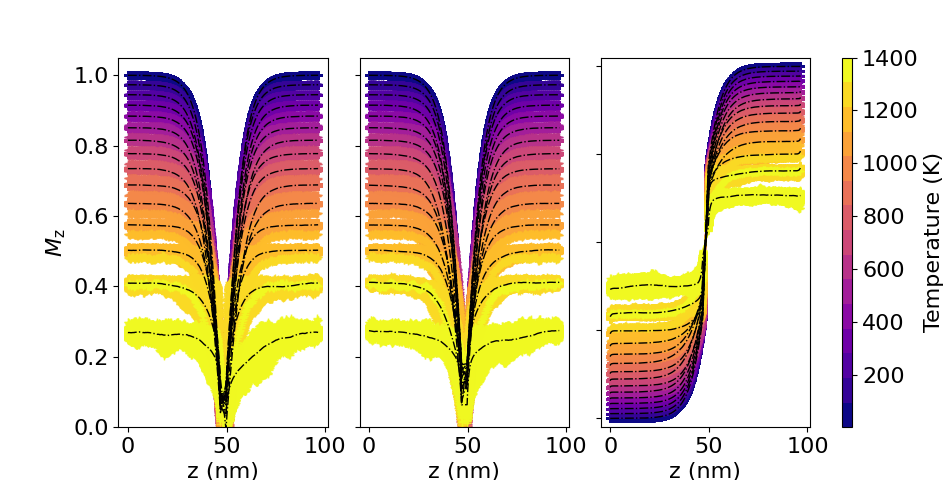}
	\caption{Magnetization profiles across the Co/Ru/Co system for various temperatures and Ru spacer thicknesses. Left: Ru spacer thickness is 1 atomic layer, the coupling across the spacer is antiferromagnetic; Center: Ru spacer thickness is 2 atomic layers, the coupling across the spacer is antiferromagnetic; Right: Ru spacer thickness is 3 atomic layers, the coupling across the spacer is ferromagnetic. The dot-dashed lines highlight the average of those profiles for each specific temperature.}
	\label{fig:ASD_CoRuCo_avg_profs}
\end{figure}

Calculating domain wall properties in the Co/Ru/Co trilayer with ASD is performed analogically to that in bulk hcp Co. The initial configuration of the system is set according to sign of the interface exchange coupling. In case of 1 and 2 Ru layers, all spins are set to $\vec{S} = (0, 0, 1)$ and the system relaxes into an antiferromagnetic domain wall. In case of 3 Ru layers, spins in the left bulk are set to $\vec{S} = (0, 0, -1)$, spins in the right bulk are set to $\vec{S} = (0, 0, 1)$, and the system relaxes into a conventional domain wall. The calculation is set to run for 800,000 steps. 
The resulting averaged magnetization profiles are shown in Fig.~\ref{fig:ASD_CoRuCo_avg_profs}. The left most image shows the magnetization profile for one monolayer of Ru atoms separating the two Co slabs. The exchange coupling across the Ru spacer is antiferromagnetic. For two monolayers Ru the antiferromagentic coupling strength is reduced by about a factor of two. The corresponding magnetization profile is shown in the center of Fig.~\ref{fig:ASD_CoRuCo_avg_profs}. For three monolayers of Ru atoms, the coupling is ferromagnetic. The domain wall, which shows a jump of the magnetization at the interface, is given in the right most image.

\subsection{\label{sec:microproperties} Intrinsic properties for micromagnetics}

The acquisition of all temperature dependent intrinsic material properties of hcp Co bulk follows the temperature dependence of the saturation magnetization $M_\mathrm{s}$. 

The ASD method was used to calculate temperature dependence of magnetocrystalline anisotropy $K(T)$. We initially start with all spins pointing into the $-z$ direction. Then we apply a positive external field and integrate the stochastic LLG equation. The field is chosen such that the magnetization reverses during the simulation time. From sampling the anisotropy energy during magnetization reversal, we obtain $E_\mathrm{anis}(M_z)$. For a Stoner-Wohlfarth particle this energy curve is a parabola \cite{lu2007domain}, its maximum corresponds to the anisotropy constant $K$ times the particle volume $V$. The Stoner-Wohlfarth switching field is $H_\mathrm{sw} = 2 K/(\mu_0 M_\mathrm{s})$.

The simulation domain consists of $10 \times 10 \times 10$ and the initial configuration is ferromagnetic (all spins $\vec{S} = (0, 0, -1)$). An external magnetic field $\vec{H}_{\mathrm{ext}}$ in the $+z$ direction is turned on and the ferromagnetic configuration slowly goes from $M_z = -1$ to $M_z = 1$. For sampling $E_\mathrm{anis}(M_z)$ during magnetization reversal we do not need to apply an external field which is sufficiently large to cause switching during the simulation time. Here we set $H_{\mathrm{ext}}$ a value slightly larger than an estimate of the Stoner-Wohlfarth switching field. We approximate $M_\mathrm{s}(T)$ with the Bloch's law
\begin{equation}
	\label{eq:Msest}
	M_\mathrm{s}^\mathrm{est}(T) = M_\mathrm{s}(0)\left(1-\frac{T}{T_\mathrm{C}}\right)^{3/2},
\end{equation}     
where $T_\mathrm{C}$ is the Curie temperature. For the temperature dependence of the anisotropy we assume the Callen-Callen law \cite{callen1966present}
\begin{equation}
  \label{eq:Kest}
  K^\mathrm{est}(T) = K(0)\left( \frac{M_\mathrm{s}(T)}{M_\mathrm{s}(0)} \right)^3
\end{equation}
We use equations (\ref{eq:Msest}) and (\ref{eq:Kest}) to approximate the Stoner-Wohlfarth switching field which is used as value for the external field for simulating magnetization reversal. We validated that varying the external magnetic field up to several times $H_\mathrm{sw}$ has negligible effect on the resulting value of anisotropy $K(T)$ which supports this approach as a valid source of temperature-dependent anisotropy, instead of having to calculate the whole hysteresis curve or employing constrained Monte Carlo techniques \cite{Asselin2010}.
\begin{figure}[h]
	\centering
	\includegraphics[width=\columnwidth]{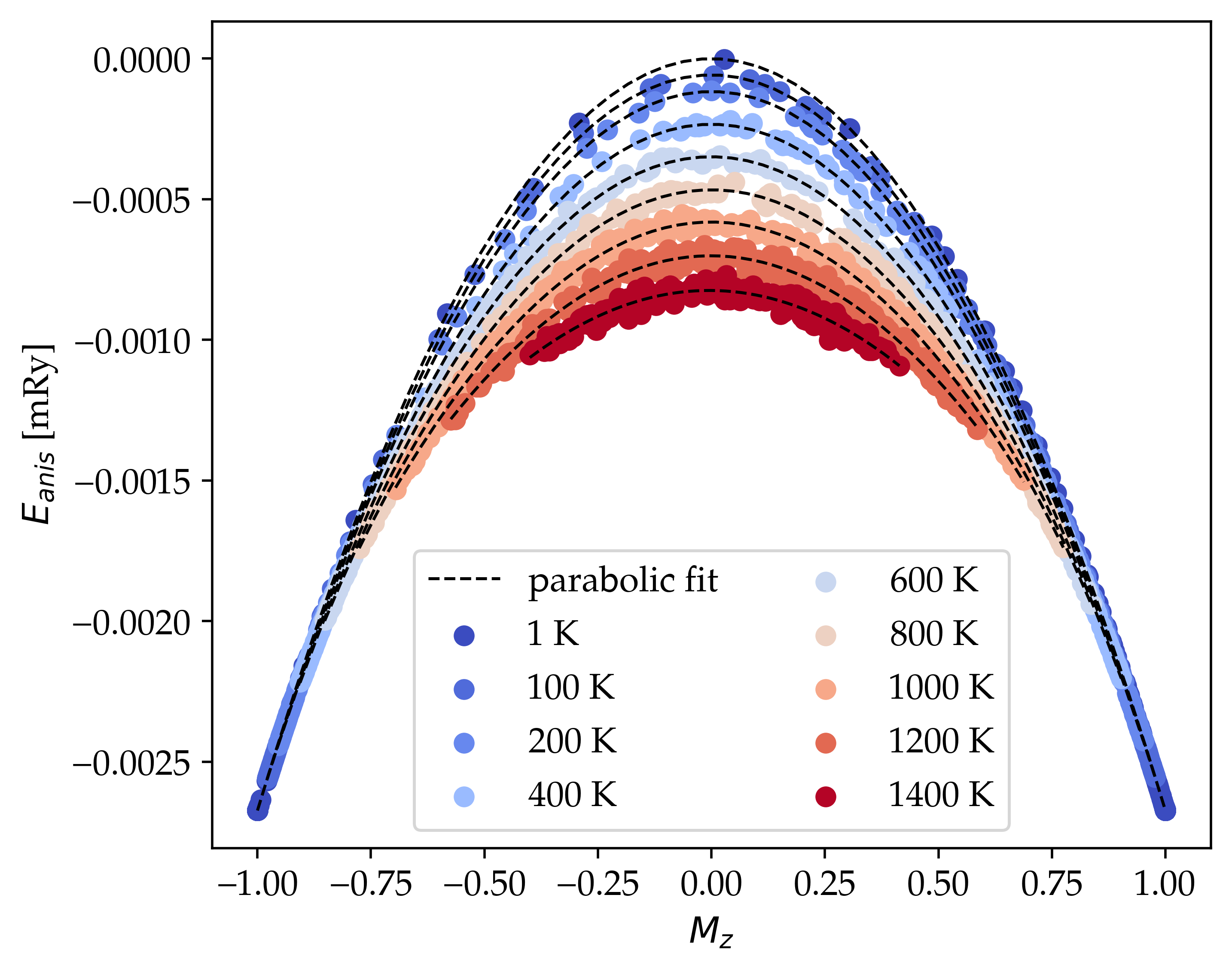}
	\caption{Energy barrier between magnetization state $M_\mathrm{z} = +1$ and $M_\mathrm{z} = -1$. Achieved through sampling of anisotropy energy plotted against magnetization in the z-direction for various temperatures in bulk hcp Co. Each dot represents an average of 50,000 ASD steps.}
	\label{fig:ASD_hcpCo_Eanis_vs_mag}
\end{figure}
The anisotropy energy value is extracted from sampling the anisotropy energy component during the ASD calculation. This energy is then plotted against 
magnetization z-component ($E_{\mathrm{anis}}$ vs $M_z$) in Fig. \ref{fig:ASD_hcpCo_Eanis_vs_mag}. We can clearly see the energy barrier between $M_z = -1$ and $M_z = +1$ which corresponds to $K(T)V$. Furthermore, the data can be accurately fitted using a parabola $E_{\mathrm{anis}} = q M_z^2 + k$, illustrating consistency with the Stoner-Wohlfart model. The anisotropy energy was calculated as difference between highest and lowest energy in Fig.~\ref{fig:ASD_hcpCo_Eanis_vs_mag}: $KV = \mathrm{max}(E_{\mathrm{anis}}) - \mathrm{min}(E_{\mathrm{anis}})$.

\begin{figure}[h]
	\centering
	\includegraphics[width=\columnwidth]{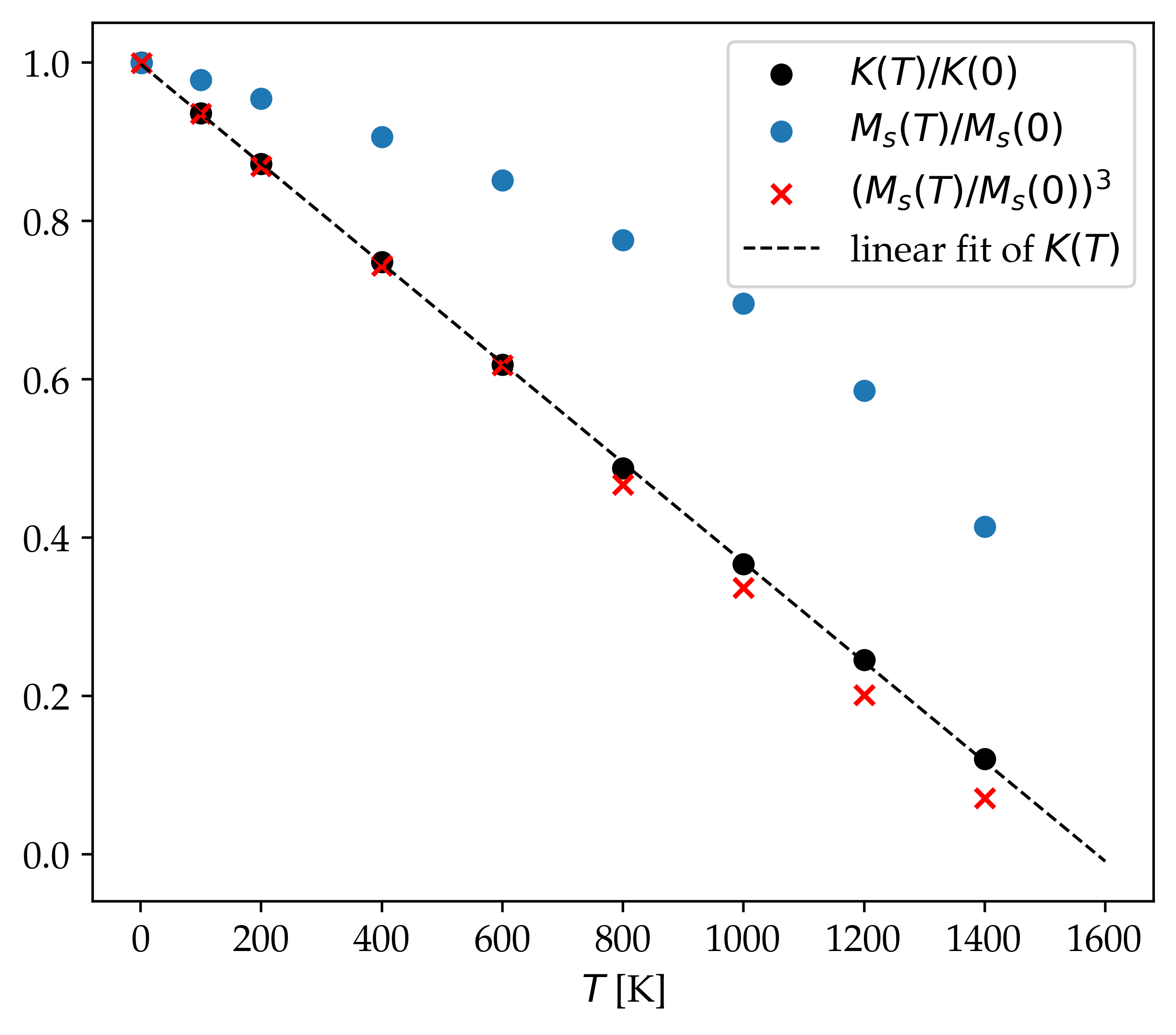}
	\caption{Temperature dependence of calculated anisotropy, saturation magnetization, and its third power in bulk hcp Co. Saturation magnetization (blue dots), spin-dynamics estimated anisotropy constant with energy barriers (black dots), anisotropy constant following Callen-Callens law (red crosses) and a linear fit (dashed line).}
	\label{fig:ASD_hcpCo_anis_vs_temp}
\end{figure}

Fig. \ref{fig:ASD_hcpCo_anis_vs_temp} shows the calculated temperature dependence of $K(T)$. The anisotropy constant decreases linearly with temperature up to about $1600~\mathrm{K}$ where it goes to zero. At low temperatures, it agrees exceptionally well with the Callen-Callen law. Therefore, we use equation (\ref{eq:Kest}) for $K(T)$ in the micromagnetic simulations. 

In order to obtain the exchange constant $A(T)$, we assume a general power law \begin{equation}
A_\beta(T) = A(0) \left( \frac{M_\mathrm{s}(T)}{M_\mathrm{s}(0)} \right)^\beta
\end{equation}
and minimize the squared error in the domain wall width
\begin{equation}
	\label{eq:ferror}
	F(\beta) = \sum_i \left( \delta_\mathrm{DW}(T_i) - \pi \sqrt{\frac{A_\beta(T_i)} {K(T_i)}}  \right)^2.
\end{equation}
In (\ref{eq:ferror}) the sum is over all temperature $T_i$ used in the ASD simulations of the domain wall width shown (see Fig.~\ref{fig:ASD_hcpCo_width_satmag}). For minimizing (\ref{eq:ferror}), we used the adaptive nonlinear least-squares algorithm method (NL2SOL) by Dennis et al. \cite{dennis1981algorithm} as implemented in the Dakota suite \cite{adams2020dakota}. From nonlinear least-square fit we obtained $\beta = 2$.   

\begin{figure}[h] 
	\includegraphics[width=\columnwidth]{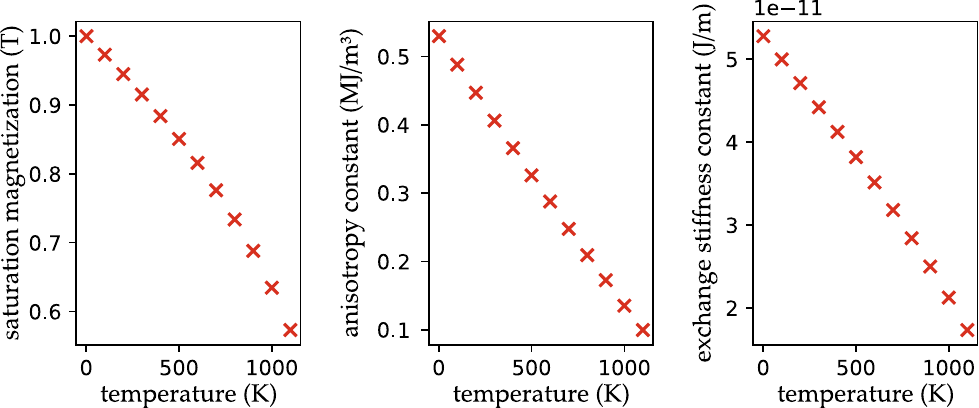}
	\caption{\label{fig:bulkcoprops}Intrinsic material properties derived from the ASD simulations. Left: Reduced temperature dependent magnetization $M_\mathrm{s}(T)/M_\mathrm{s}(0)$. Center: Anisotropy constant $K(T)$. Right: Exchange stiffness constant $A(T)$.}
\end{figure}

Fig.~\ref{fig:bulkcoprops} shows the temperature dependent intrinsic material properties for hcp Co.

In addition to the bulk properties of hcp Co, we need the coupling constant $J_\mathrm{Co,Co}(T)$ across the Ru layer. Given the angle of the magnetic moments between the interfaces $\varphi_\mathrm{int}$ and with the knowledge about bulk exchange constant $A(T)$ and magnetocrystalline anisotropy $K(T)$ at specific temperature $T$ we can use equation (\ref{eq:JCoCodmz}) to obtain the temperature coupling constant from the ASD simulations. 

We start from the ASD simulation of the magnetization profiles across the Co/Ru/Co trilayer presented in section \ref{sec:spindynamicprofiles} and compute the average magnetic moments of the Co atoms left and right of the interface. Using the notation of section \ref{sec:spindynintro}, the average magnetic moments are
\begin{eqnarray}
	\mathbf{m}^\mathrm{spindyn}_\mathrm{left} &=& \frac{1}{|C_{IL}|} \sum_{i \in C_{IL}} \vec{S}_i \\
    \mathbf{m}^\mathrm{spindyn}_\mathrm{right} &=& \frac{1}{|C_{IR}|} \sum_{i \in C_{IR}} \vec{S}_i
\end{eqnarray} 
where $|C_{IL}|$ and $|C_{IR}|$ are the number of Co atoms in the sets of atoms to the left and to the right of the interface, respectively. We now compute the angle $\varphi_\mathrm{int}$ between the two unit vectors $\mathbf{m}^\mathrm{spindyn}_\mathrm{left}$ and $\mathbf{m}^\mathrm{spindyn}_\mathrm{right}$.

\begin{figure}[htb]
	\includegraphics[width=0.65\columnwidth]{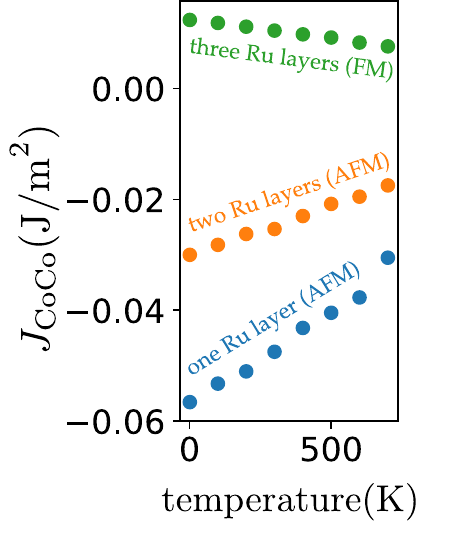}
	\caption{\label{fig:dww-angle} The interface exchange coupling extracted from the $\Delta m_{z}$ relation from the Eq.~\ref{eq:JCoCodmz} in units of J/m$^2$.}
\end{figure}

Fig.~\ref{fig:dww-angle} shows the resulting coupling constant $J_\mathrm{Co,Co}(T)$ for one, two, and three monolayers of Ru atoms between the Co layers as a function of temperature $T$. 

In Tab.~\ref{tab:JCoCo} we show selected values for the micromagnetic input parameters.

\begin{table}[h]
\caption{Input parameters for micromagnetic simulations of Co/Ru/Co trilayers. The columns give the number Ru monolayers $n$, the temperature $T$, exchange constant $A$, the anisotropy constant $K$ and the coupling constant $J_{Co,Co}$.}
\label{tab:JCoCo}
\begin{tabular}{c r c c c}
$n$ & $T (\mathrm{K})$ & $A (\mathrm{pJ/m})$ & $K (\mathrm{MJ/m}^3)$ & $J_\mathrm{Co,Co} (\mathrm{J/m}^2$)\\ \hline 
1 & 1   & 52.78 & 0.547 & $-0.05657$ \\
2 & 1   & 52.78 & 0.547 & $-0.03000$ \\
3 & 1   & 52.78 & 0.547 & $ 0.01236$ \\
1 & 100 & 50.29 & 0.509 &$-0.053234$ \\
2 & 100 & 50.29 & 0.509 & $-0.02820$ \\
3 & 100 & 50.29 & 0.509 &  $0.01182$ \\
1 & 300 & 45.04  & 0.431 & $-0.04749$ \\
2 & 300 & 45.04 & 0.431 & $-0.02535$ \\
3 & 300 & 45.04 & 0.431 &  $0.01044$ \\
1 & 500 & 39.53 & 0.355 & $-0.04046$ \\
2 & 500 & 39.53 & 0.355 & $-0.02082$ \\
3 & 500 & 39.53 & 0.355 &  $0.00919$ \\
\hline
\end{tabular}
\end{table} 

\subsection{\label{sec:micromagneticprofiles}Domain wall profiles from micromagnetics}

We now compute magnetization profiles across Co/Ru/Co layers using micromagnetic simulations. For the simulations we use the temperature dependent properties listed in Tab.~\ref{tab:JCoCo} as input. The wall profiles are obtained by minimizing the micromagnetic energy (\ref{eq:mumagenergy}) numerically. 

\begin{figure}[h] 
    \includegraphics[width=\columnwidth]{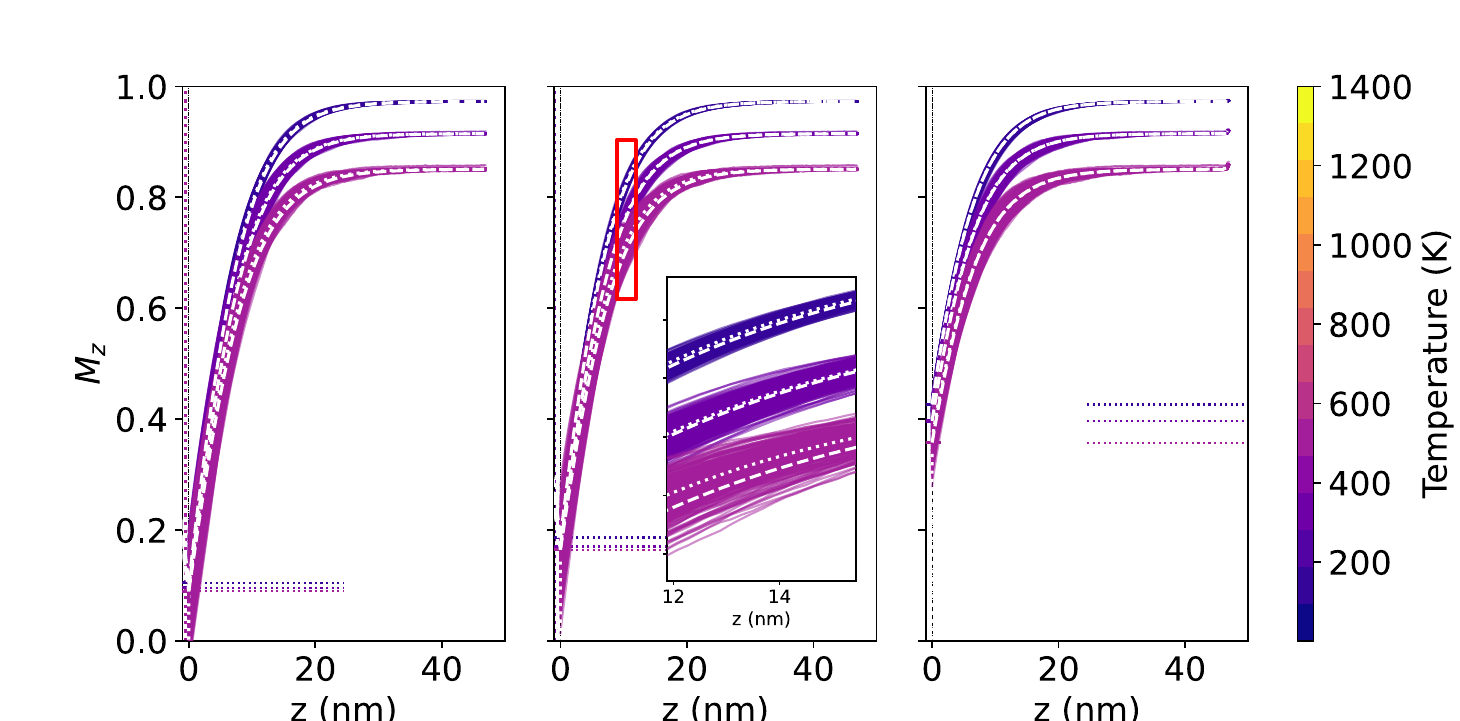}
	\caption{\label{fig:compare}Comparison of temperature dependent magnetization profiles obtained from the ASD and algebraic solution. The colored surfaces show the ensemble of all spin-dynamics simulations. The white dotted lines show the average values (see Fig.~\ref{fig:ASD_CoRuCo_avg_profs}), the dashed white line demonstrate the results from algebraic solution. Left and Mid: Antiferromagnetic coupling across one and two monolayers Ru. Right: Ferromagnetic coupling across three monolayer Ru atoms. Note: We show only the right side of the profiles because of symmetry.}
\end{figure}

\section{\label{sec:conclusion}Conclusion}

In lowest order, temperature is included in micromagnetism through temperature-dependent material parameters \cite{skomski2013finite}. By passing zero-temperature parameters such as the exchange integrals obtained from \emph{ab initio} calculations to the ASD simulations, temperature dependent values for the spontaneous magnetization, the exchange constant, and the anisotropy constant can be obtained \cite{atxitia2010multiscale,nieves2017atomistic,westmoreland2018multiscale}. In addition to these bulk material properties, the interface properties are essential for treating magnetization processes in multimaterial and multilayered systems. A prominent example is the coupling of ferromagnetic materials through a thin Ru layer in Co/Ru/Co multilayers. Depending on the thickness of Ru spacer, the coupling may be ferromagnetic or antiferromagnetic \cite{parkin1990oscillations}. Such interfaces can be described within micromagnetic theory if the energy is augmented with interface energy terms. This interface energy term leads to a jump of the magnetization across the interface. The coupling constant between the magnetization vectors at both sides of the interface will depend on temperature. It can be derived from the domain wall energy in the bulk and the size of the jump of the magnetization at the interface. The latter can be computed from spin dynamic simulations. 

Our work provides a systematic approach to derive both the temperature-dependent bulk properties and the temperature-dependent coupling constant for magnetic multilayers combining \emph{ab initio} calculations, spin dynamics simulations, and micromagnetics.

\section{Acknowledgments}
We gratefully acknowledge the financial support by the Austrian Science Fund (FWF) I 5712. This work was supported by the project e-INFRA CZ $(ID:90254)$ for CPU time resources and the QM4ST project $(CZ.02.01.01/00/22\_008/0004572)$ by the Ministry of Education, Youth and Sports of the Czech Republic and by Grant No. $22-35410K$ by the Czech Science Foundation.

\bibliography{MAGINT}

\end{document}